# The envelope of projectile trajectories in midair

Peter S  Chudinov

Department of Theoretical Mechanics, Faculty of Applied
Mathematics and Physics, Moscow Aviation Institute, Russia

E-mail: choudin@k804.mainet.msk.su

## Abstract

A classic problem of the motion of a point mass (projectile) thrown at an angle to the horizon is reviewed. The air drag force is taken into account with the drag factor assumed to be constant. Analytic approach is used for investigation. Simple analytical formulas are used for the constructing the envelope of the family of the point mass trajectories. The equation of envelope is applied for determination of maximum range of flight. The motion of a baseball is presented as an example.

## 1. INTRODUCTION

The problem of the motion of a point mass thrown at an angle to the horizon is a constituent of many introductory courses on physics. With zero air drag force, the analytic solution is well known. The trajectory of the point mass is a parabola.  With air drag taken into account, a finding the main variables of the problem is reduced to quadratures [1 – 4].  Appropriate integrals are not taken in finite form. The problem, to all appearance, does not have the exact analytic solution, and therefore in most cases is solved numerically [5]. Analytic approaches to the solution of the problem are not sufficiently advanced. Meanwhile, analytic solutions are very convenient for a straightforward adaptation to applied problems and especially useful for a qualitative analysis. Comparatively simple approximate analytical formulas to study the motion of the point mass in a medium with quadratic drag force have been obtained  using such an approach  [6 – 7] . These formulas make it possible to carry out a complete qualitative analysis without using numeric integration of point mass motion differential equations.

This paper considers application of formulas [6 – 7] for the constructing the envelope of the family of the point mass trajectories.  Family of trajectories is formed when throwing a point mass with one and same initial velocity, but with different angles of throwing. Problem of constructing the envelope is solved by means of simple analytical formulas. Equation of the envelope is used for finding of maximum range of flight of point mass in the case when the spot of incidence is above or below than spot of throwing. In the examples cited is shown that obtained formulas ensure high accuracy of determination of maximum range. Mistake of analytical calculation of range does not exceed 1%.

## 2. EQUATIONS OF POINT MASS MOTION AND ANALYTICAL FORMULAS FOR BASIC PARAMETERS

The problem of the motion of a point mass in air, with a number of conventional assumptions, in case of the drag force proportional to the square of the velocity, $R = mgkV^2$, boils down to numerical integration of the differential system [4]

$$\frac{dV}{dt} = -g\sin\theta - gkV^2, \quad \frac{d\theta}{dt} = -\frac{g\cos\theta}{V}, \quad \frac{dx}{dt} = V\cos\theta, \quad \frac{dy}{dt} = V\sin\theta. \quad (1)$$

Here $V$ is the velocity of the point mass, $\theta$ is the angle between the tangent to the trajectory of the point mass and the horizontal, $g$ is the acceleration due to gravity, $m$ is the mass of the particle, $x, y$ are the Cartesian coordinates of the point mass, $k = \frac{\rho_a c_d S}{2mg} = const$ is the proportionality factor, $\rho_a$ is the air density, $c_d$ is the drag factor for a sphere, and $S$ is the cross-section area of the object.

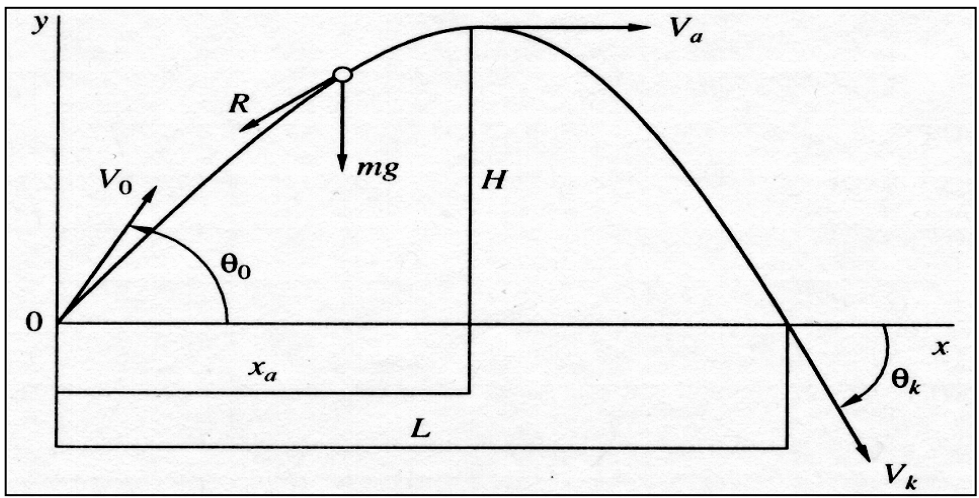

Figure 1. Basic motion parameters.

Comparatively simple approximate analytical formulas for the eight basic parameters of point mass motion are obtained in [6 – 7]. Let us set out the relationships required for the maximum ascent height $H$, velocity at the trajectory apex $V_a$, motion time $T$, flight range $L$, and the abscissa of the trajectory apex $x_a$ (Figure 1):

$$H = \frac{V_0^2 \sin^2 \theta_0}{g(2 + kV_0^2 \sin \theta_0)}, \quad V_a = \frac{V_0 \cos \theta_0}{\sqrt{1 + kV_0^2 \left(\sin \theta_0 + \cos^2 \theta_0 \ln \tan\left(\frac{\theta_0}{2} + \frac{\pi}{4}\right)\right)}},$$

$$T = 2\sqrt{\frac{2H}{g}}, \quad L = V_a T, \quad x_a = \sqrt{LH \cdot \cot \theta_0}. \quad (2)$$

With zero drag ($k = 0$), formulas (2) go over into respective formulas of the point mass parabolic motion theory. All motion characteristics described by relationships (2) are functions of $V_0, \theta_0$, initial conditions of throwing. Relationships (2), in turn, make it possible to obtain simple analytical formulas for six basic functional relationships of the problem which are $y(x)$, $y(t)$, $x(t)$, $y(\theta)$, $x(\theta)$, $t(\theta)$ [6].

One of the most important aspects of the problem is determination of an optimum angle of throwing of a point mass which provides the maximum range. The equation for the optimum angle of throwing $\alpha$ in the case when the points of incidence and throwing are on the same horizontal is obtained in [7]:

$$\tan^2 \alpha + \frac{p \sin \alpha}{4 + 4p \sin \alpha} = \frac{1 + p\lambda}{1 + p(\sin \alpha + \lambda \cos^2 \alpha)}. \quad (3)$$

Here $p = kV_0^2$, $\lambda(\alpha) = \ln \tan\left(\frac{\alpha}{2} + \frac{\pi}{4}\right)$. We use formulas (2) - (3) for the constructing the envelope.

## 3. THE EQUATION OF THE ENVELOPE IN MIDAIR

In the case of no drag the trajectory of a point mass is a parabola. For the different angles of throwing under one and same initial velocity projectile trajectories form a family of parabolas. Maximum range and maximum height for limiting parabolas are given by formulas

$$L_{max} = \frac{V_0^2}{g}, \quad H_{max} = \frac{V_0^2}{2g}. \quad (4)$$

The envelope of this family is also a parabola, equation of which is usually written as [8]

$$y(x) = \frac{V_0^2}{2g} - \frac{g}{2V_0^2} x^2. \quad (5)$$

Using (4), we will convert the equation (5) as

$$y(x) = \frac{H_{max}(L_{max}^2 - x^2)}{L_{max}^2}. \quad (6)$$

We will set up an analytical formula similar to (6) for the envelope of the point mass trajectories taking into account the air drag force.

Taking into account the formula (6), we will derive an equation of the envelope as

$$y(x) = \frac{H_{max}\left(L_{max}^2 - x^2\right)}{L_{max}^2 - ax^2}. \tag{7}$$

Such structure of equation takes into account the fact that the envelope has a maximum under $x = 0$. Besides, function (7) under $a > 0$ has a vertical asymptote, as well as any point mass trajectory accounting resistance of air. In formula (7) $H_{max}$ is the maximum height, reached by the point mass when throwing with initial conditions $V_0$, $\theta_0 = 90°$; $L_{max}$ - the maximum range, reached when throwing a point mass with the initial velocity $V_0$ under some optimum angle $\theta_0 = \alpha$. In the parabolic theory an angle $\alpha$ is $45°$ under any initial velocity $V_0$. Taking into account the resistance of air, an optimum angle of throwing $\alpha$ is less than $45°$ and depends on the value of parameter $p = kV_0^2$. A choice of positive factor $a$ in the formula (7) is sufficiently free. However it must satisfy the condition $a = 0$ in the absence of resistance ($k = 0$). We will define this factor by formula

$$a = 2 - \sqrt[3]{2\left(\frac{L_{max}}{x_a}\right)^2}. \tag{8}$$

The parameter $x_a$ in (8) is an abscissa of the top of trajectory that has the maximum range. At the $k = 0$ a ratio of parameters $L_{max}/x_a = 2$, and factor $a$ vanishes. Thereby, according to the formula (7), three parameters are required for the construction of an envelope: $H_{max}$, $L_{max}$, $x_a$. Parameter $H_{max}$ with the preceding notation is defined by formula [4]

$$H_{max} = \frac{1}{2gk}\ln\left(1 + kV_0^2\right). \tag{9}$$

We will calculate parameters $L_{max}$, $x_a$ as follows. Under a given value of quantity $p = kV_0^2$ we will find the root $\alpha$ of equation (3). An angle $\alpha$ ensures the maximum range of the flight. Substituting value $\alpha$ in formulas (2), we define the required parameters:

$$L_{max} = L(\alpha) = V_a(\alpha)\cdot T(\alpha), \qquad x_a = \sqrt{L_{max}H(\alpha)\cdot\cot\alpha}. \tag{10}$$

The equation of the envelope can be used for the determination of maximum range if the spot of falling lies above or below the spot of throwing. Let the spot of falling be on a horizontal straight line defined by the equation

$y = y_1 = const$. We will substitute a value $y_1$ in the equation (7) and solve it for $x$. We obtain the formula

$$x_{max} = L_{max} \sqrt{\frac{H_{max} - y_1}{H_{max} - ay_1}}. \qquad (11)$$

The correlation (11) allows us to find a maximum range under the given height of the spot of falling.

## 4. THE RESULTS OF THE CALCULATIONS

As an example we will consider the moving of a baseball with the resistance factor $k = 0.000548$ s$^2$/m$^2$ [5]. Other parameters of motion are given by values

$$g = 9.81 \text{ м/с}^2, \ V_0 = 45 \text{ m/s}, \ y_1 = \pm 20 \text{ m}, \ \pm 40 \text{ m}.$$

Substituting values $k$ and $V_0$ in the formula (9), we get $H_{max} = 69.4$ m. Hereinafter we solve an equation (3) at the value of non-dimensional parameter $p = kV_0^2 = 1.11$. The root of this equation gives the value of an optimum angle of throwing. This angle ensures the maximum range: $\alpha = 40.6°$. Substituting this value in formulas (10), we have:

$$L_{max} = 117.4 \text{ m}, \ x_a = 66.3 \text{ m}.$$

According to the formula (8) the factor $a = 0.168$. The graph of the envelope (7) is plotted on Figure 2 together with the family of trajectories. We note that family of trajectories is received by means of numerical integrating of the equations of motion of a point mass (1). A standard fourth-order Runge-Kutta method was used.

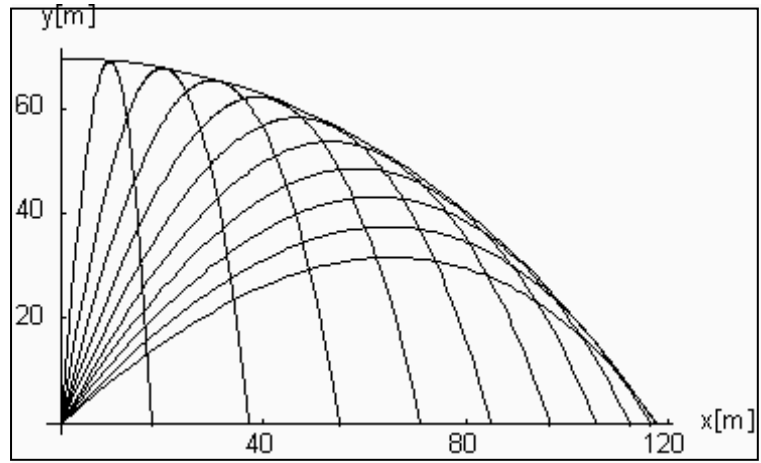

Figure 2. The family of projectile trajectories and the envelope of this family.

The results of calculations using the formula (11) are presented in the Table 1.

TABLE 1. MAXIMUM RANGE UNDER DIFFERENT HEIGHTS OF THE SPOT OF THE FALLING

| Value $y_1$ (m) | Parameter (m) | Analytical value | Numerical value | Error ( % ) |
|---|---|---|---|---|
| 40 | $x_{max}$ | 80.4 | 79.8 | 0.8 |
| 20 | $x_{max}$ | 101.5 | 101.4 | 0.1 |
| 0 | $L_{max}$ | 117.4 | 117.8 | –0.4 |
| –20 | $x_{max}$ | 130.1 | 131.1 | – 0.8 |
| –40 | $x_{max}$ | 140.7 | 142.3 | – 1.1 |

Numerical values of the range in the table 1 are also received by integrating a system of equations of motion (1). The tabulated data show that formulas (2), (3), (7), (11) ensure sufficiently pinpoint accuracy of calculation of parameters of motion.

## 5. CONCLUSION

The proposed approach based on the use of analytic formulas makes it possible to significantly simplify a qualitative analysis of the motion of a point mass with the air drag taken into account. All basic parameters of motion, functional relationships and various problems of optimization are described by simple analytical formulas. Moreover, numerical values of the sought variables are determined with an acceptable accuracy. Thus, many formulas [6] – [7] and (2) – (11) make it possible to carry out complete analytical investigation of the motion of a point mass in a media with drag same as it is done for the case of no drag.